\documentclass[12pt]{article}
\date{}

\title{Free-Surface Hydrodynamics in Conformal Variables: Are Equations of Free-Surface Hydrodynamics on Deep Water Integrable?}

\author{V. Zakharov\\
{\small Department of Mathematics, University of Arizona, Tucson, USA;}\\
{\small Novosibirsk State University, Novosibirsk, Russia;}}
\vspace{-0.5cm}
\begin{document}
\maketitle

\begin{abstract}

The hypothesis on complete integrability of equations describing the potential motion of incompressible ideal fluid with free surface in 2-D space in presence and absence of gravity was formulated by Dyachenko and Zakharov in 1994 [1]. Later on the same authors found that these equations have indefinite number of additional motion constants [2] that was an argument in support of the integrability hypothesis. In this article we formulate another argument in favor of this conjecture. It is known [3] that the free-surface equations have an exotic solution that keeps the surface flat but describes the compression of the whole mass of fluid. In this article we show that the free-surface hydrodynamic is integrable if the motion can be treated as a finite amplitude perturbation of the compressed fluid solution. Integrability makes possible to construct an indefinite number of exact solutions of the Euler equations with free surface.

\end{abstract}

\section{Basic equations}

We study the potential flow of two-dimensional ideal incompressible fluid.
The fluid occupies a half-infinite domain
$$
-\infty < y < \eta(x,t), \,\,\,\,\,\, -\infty < x < \infty.
$$
The flow is potential, so that
$v = \nabla \Phi, \,\,\,
\Phi\vert_{y=\eta(x,t)} = \psi(x,t)$.
Boundary conditions on the surface are standard.
It is known that the shape of surface $\eta(x,t)$
and the
potential on the surface $\psi(x,t)$ form a pair of canonically conjugated
variables obeying the Hamiltonian equations:
$$
\frac{\partial \eta}{\partial t} = \frac{\delta {\cal H}}{\delta \psi}, \qquad
\frac{\partial \psi}{\partial t} = -\frac{\delta{ \cal H}}{\delta \eta}.
$$
Here ${\cal H}$ is Hamiltonian function, the total energy of the fluid [4].

These equations  minimize the action
$$
S = \int L dt, \qquad
L = \int_{-\infty}^{\infty} \psi\eta_t dx -{\cal H}.
$$
Starting from this point let us forget for a while about hydrodynamics, and
consider more general case. Namely,
let's think of $\cal H$ as some arbitrary functional of $\psi$ and $\eta$.

Let $z(w,t)$ be the conformal mapping of the domain,
bounded by the curve $\eta(x,t)$
to the lower half-plane of $w$
$$
w = u+ iv, \qquad -\infty <u <\infty, \quad -\infty <v <0
$$
Hamiltonian function $\cal H$
can be considered as a functional, depending only on $\psi$ and $y$.
Hence, one can put $\frac{\delta\cal H}{\delta x} = 0$.

The condition $\delta S = 0$ leads to the following ``implicit''
equations of motion
$$
y_t x_u-x_t y_u = \frac{\delta\cal H}{\delta\psi}
$$
$$
\psi_t x_u -x_t\psi_u -\hat H(y_u\psi_t-y_t\psi_u) =
-\frac{\delta\cal H}{\delta y }
$$
Thereafter we use the Hilbert transformation:
$$
\hat H f=\frac{1}{\pi} \int \frac{f(s)}{s-w}\,ds
$$
Later on we denote:
$$
z=x+iy
$$
$$
\Phi=\Psi +i \hat H \Psi
$$
These complex-valued functions are analytic in the lower half-plane $v\leq 0$.

Equations for "implicit" equations of motion can be rewritten as follows [5]:
$$
z_t \bar{z}_u-\bar{z}_t z_u=-\Phi_u +\bar{\Phi}_u
$$
$$
\Psi_t z_u -\Psi_u z_t + \frac{1}{2}\frac{\bar{\Phi}_u^2}{\bar{z}_u}=0
$$

\section{Self-similar solutions}

Equations
$$
z_t \bar{z}_u-\bar{z}_t z_u=-\Phi_u +\bar{\Phi}_u
$$
$$
\Psi_t z_u -\Psi_u z_t + \frac{1}{2}\frac{\bar{\Phi}_u^2}{\bar{z}_u}=0
$$
admit the following substitution [6]:
$$
z=t^{\alpha} z_0(u)
$$
$$
\Phi=t^{2\alpha -1}  \Phi_0 (u)
$$
Then, the self-similar solutions are
$$
\eta=t^{\alpha} F(\frac{x}{t^{\alpha}}), \qquad t\to t-t_0
$$
In the presence of gravity only one solution is possible, $\alpha=2$:
$$
\eta=g(t_0 -t)^2 F(\frac{x}{g (t_0 -t)^2})
$$
This is formation of wedge with $\alpha=120^0$. If $g=0$, all $\alpha$ are possible:
$$
\alpha(z_0 \bar{z}_{0u})=\bar{\Phi}_{0u}-\Phi_{0u}
$$
$$
(2\alpha -1) \Psi_0 z_{0u} -\alpha \Psi_{0u} z +\frac{1}{2} \frac{\bar{\Phi}_u^2}{\bar{z}_u}
$$
$$
\Psi_0=\frac{1}{2}(\Phi_0 +\bar{\Phi}_0)=Re  \Phi_0
$$

If $\alpha=-3$, there is a parabolic Dirichlet jet. 
If $\alpha=-1$, there is a compressed fluid.

\section{Self-similar compressed fluid}

Longuet-Higgins [3] found the following solution of the Euler equations [3]:
$$
\eta\equiv 0
$$
$$
\Phi(x,y,t)=\frac{1}{2}\frac{1}{t-t_0} (x^2 -y^2)
$$
$$
P=-\frac{y^2}{(t-t_0)^2} \qquad P=0, y=0
$$
In conformal variables we have:
$$
z_0=tu \qquad \Phi_0=\frac{1}{2} tu^2
$$
Then equations for the shape of self-similar solutions are satisfied. Let us study perturbation of this solution:
$$
z\to ut+z \qquad \Phi\to \frac{1}{2} u^2 t + \Phi
$$
Equations for the self-similar solutions read
$$
tz_t -uz_u +\Phi_u=P^-(\bar{z}_t z_u -z_t \bar{z}_u)
$$
$$
P^-\left\{ \frac{u}{2} (uz_u -\Phi_u)+t(\frac{1}{2} \Phi_t -uz_t) +\Psi_t z_u -\Psi_u z_t\right\}=0
$$

\underline{Miracle number 1}

These solutions are satisfied if
$$
z=\alpha(u)\qquad \Phi=\Phi_0 (u)= \partial^{-1} u \alpha (u)
$$
Function $\alpha(u)$ is an \underline{arbitrary!} function analytic in the lower half-plane
$$
\alpha(w)\to 0\qquad Im w \to -\infty
$$

Let
$$
\alpha=\frac{A}{u+ia} \qquad A,a - real \,\,\,constants, u>0
$$
Shape of the surface is presented in the parametric form
$$
x=u+\frac{Aut}{u^2 +a^2 t^2} \qquad y=-\frac{aAt^2}{u^2 +a^2 t^2}
$$
$$
\frac{\partial x}{\partial u}\to 1 \qquad at\qquad t\to\pm \infty
$$
Bifurcation condition $\partial z/\partial u =0$ leads to the following expression:
$$
u^2=\frac{1}{2} A \,t \Big( 1\pm\sqrt{1-\frac{8a^2}{A^2}}\Big) -a^2 t^2
$$
If
$
a^2>\frac{1}{8} A^2
$
the solution is one-valued. If
$
a^2<\frac{1}{8} A^2
$
ie, the pole is close to the real axis, we obtain invertible:

1. Formation of bubbles (if $A>0$)

2. Formation of droplets (if $A<0$)

 \underline{Miracle number 2}

Let us look for solution of the above equations in the form:
$$
z=\alpha(u)+\frac{1}{t}\, z_1(u)+\frac{1}{t^2}\, z_2 (u)+\cdots
$$
$$
\Phi=\Phi_0 (u) +\frac{1}{t}\,\Phi_1(u)+\frac{1}{t^2}\,\Phi_2(u)+\cdots
$$
Now again $z_1(u)$ is arbitrary function analytic in the lower half-plane
$$
\Phi_1(u)=u\,z_1(u)
$$
$$
u\,z_2(u)=-P^-\,(\bar{z}_1 \alpha_u -z_1\,\bar{\alpha}_u)
$$

The system is integrable!

There is another form of complex equations.
Following Dyachenko [7], we introduce new variables:
$$
R = \frac{1}{z'},\qquad
V = i\frac{\partial\Phi}{\partial z} = i R\Phi'.
$$
For the simplest case of absence of gravity the Dyachenko equations read
$$
R_t = i(UR' - RU')
$$
$$
V_t = i(UV' -R B')
$$
In $R$ and $V$ variables:
$$
U = \hat P^-(R\bar V + \bar R V),\quad
B =  \hat P^-(V\bar V)
$$

In the presence of gravity the first equation is not changed. The second one takes the form:
$$
V_t = i\Big( UV' - R \hat P^-(V\bar V)' +g(R-1) \Big)
$$

\section{Poles and cuts}

Functions $R, V, U, B$ are analytic on $Jm\,w<0$. Moreover, $R\neq 0$, $Jm\,w<0$.
However these functions may have
singularities on upper half-plane. Function $R$ can have zeros at $Jm w>0$.

The following facts are important:

1. Zeroes of $R$ (denote them $\lambda_n$) are persistent: $R(\lambda_n)=0$. They cannot appear or disappear and move obeying the law
$$
\dot\lambda_n=i\,U_n,\qquad U_n=U|_{w=\lambda_n}
$$

2. Cuts are persistent if they are of root square type.
If in the initial moment of time
$$
R=1+\frac{\gamma}{w-a},\qquad Jm (a-\gamma)>0
$$
the pole is non-persistent and turns to the cut immediately.

Let $w=w_n$ be a branch point. In the neighborhood of these branch points
$$
R=R_1(w-w_n)^{1/2} \qquad R|_{w=w_n}=0
$$
Moreover
$$
\dot w_n=i V_n \qquad u_n=U|_{w=w_n}
$$

\section{Approximation for narrow cuts}

Suppose that cuts for $R$ and $V$ are far from the real axis, namely
their width $(b-a)$ is much less then the distance to the real axis
$(b-a)\ll a$ [2]. Then one can approximate
$$U =\hat P^-(\bar V R + V\bar R)\quad as \quad
U \simeq V_c R + V R_c - V_c
$$
Here $V_c$ is the value of $\bar V$ at the some point on the narrow
cut of $R$, and $R_c$ is the value of $\bar R$ on the narrow cut of $V$.
The last term appears due to asymptotic of $R$ at infinity.

There is a ground for
this approximation. Both $\bar R$ and
$\bar V$ have singularities in the lower half-plane, at the complex
conjugate points with respect to $R$ and $V$.

If we consider
narrow cut (at the same place for $R$ and $V$), that means we assume
$V\simeq V_c$ and $R\simeq R_c$
being time dependent only.
This assumption allows us to get the approximate expression for
$B$ also:
$$
B = \hat P^-(\bar V V)\simeq V_c V.
$$
For the limiting case of infinitely
narrow cut the approximation is exact.

Substituting $U$ and $B$ into Dyachenko equations
we end up with the following equations:
$$
\dot R+iV_cR' = iR_c(VR'-V'R), \qquad
\dot V+iV_cV' = iR_c(VV')
$$
In the moving framework
$\chi = w -i\int_0^t V_c\,dt$
these equations read:
$$
\dot R = iR_c(VR'-V'R), \qquad
\dot V = iR_c(VV')
$$
Now, the space derivative is with respect to $\chi$.
It is remarkable that we derived complex Hopf's equation.

If
we introduce the new time $\tau(t)$, so that
$\dot \tau(t) = R_c(t)$,
and
$$
\chi = w -i\int_0^\tau\frac{V_c}{R_c}\,d\tau,
$$
then, recall that $R = \frac{1}{z'}$, we finally get the following set
of quadratic equations:
$$
z_{\tau} = iVz', \qquad
V_{\tau} = iVV'
$$
The solution of equation that satisfies the initial conditions is:
$$
V(\chi,\tau) = \frac{\lambda+i\chi-\sqrt{(\lambda+i\chi)^2 -4A\tau}}{2\tau}
$$
The branch of square root is chosen to provide zero
asymptotic for $V$ at infinity.
The general solution  with velocity $V(\chi,t)$ satisfying
the Hopf's equation is given by formula $
z(\chi,\tau) = G(i\chi-\tau V) $
with arbitrary function $G$. From the initial conditions one
can easily obtain that
$G(\xi) = -i\xi,$
and for $z(\chi,\tau)$ and $R(\chi,\tau)$ get the following expressions:
\begin{eqnarray}
z(\chi,\tau) &=& -\frac{i}{2}\{-\lambda +i\chi +\sqrt{(\lambda+i\chi)^2
-4A\tau}\}\cr
R(\chi,\tau) &=& \frac{2\sqrt{(\lambda+i\chi)^2 -4A\tau}}
{\lambda+i\chi+\sqrt{(\lambda+i\chi)^2 -4A\tau}}\nonumber
\end{eqnarray}
This solution describes the formation of the spray but it cannot describe the asymptotic shape of the spray at $\tau \to \infty$.

\section{Is free surface hydrodynamics integrable?}

We see that approximation of narrow cut leads to an integrable system. Is the whole system integrable? The Dyachenko equations can be rewritten in the differential form
$$
\frac{\partial}{\partial t}\,\frac{1}{R}= i \frac{\partial}{\partial w} \left(\frac{U}{R}\right) , \qquad
\frac{\partial}{\partial t}\,\frac{V}{R}= i \frac{\partial}{\partial w} \left(\frac{U V}{R}-B \right)+g\left(1-\frac{1}{R}\right)
$$
Let $I=\int_{-\infty}^{\infty} \frac{1}{R}\, du$,\,\,\,$J=\int_{-\infty}^{\infty} \frac{V}{R}\, du$. Then
$$
\frac{d I}{dt}=0,\qquad \frac{d J}{d t}=-g I,
$$
 and $I=const$, $J=J_0 -g I t$. These equalities are conservation laws of mass and horizontal component of momentum. However, these relations could be generalized.

Let $\Gamma$ be a closed contour and all functions be analytic in some neighborhood of this contour,
$$
I=\oint_{\Gamma} \frac{1}{R}\,\,d w,\qquad J=\oint_{\Gamma} \frac{V}{R}\,\,d w,
$$
and $I,\,\,J_0$ be motion constants.
If in a vicinity of $\lambda_n$, $R$ and $V$ can be presented as follow
$$
R=a_n (w-\lambda_n) +\cdots \qquad V=b_n+b_1(w-\lambda_n)+\cdots
$$
then
\begin{eqnarray}
\frac{d a_n}{d t}&=&0 \qquad  a_n=const \cr
\frac{d b_n}{dt}&=&-g a_n \qquad b_n=b_{0n} -g a_n t \nonumber
\end{eqnarray}

In other words, $a_n$, $b_{0n}$ are motion constants. We conclude that each zero of $R$ generates two complex (four real) motion constants. 
All motion integrals are in involution. They form the Abelian  Lie algebra. The question about the completeness of the  set of integrals is open yet.

All functions $R, V, U, B$ can be analytically continued to a certain  Riemann surface, and each list of this surface generates additional motion constants. 
This fact leads to the plausible conjecture that the whole set of motion constants is complete, hence the system is completely integrable.
The fact of integrability of the "compressed fluid" supports this conjecture.

The main question is the compactness. 
If the solution can be analytically continued to a compact Riemann surface, the system is integrable.
In this case the dynamics of the fluid if completely defined by evolution of poles on this surface.

Any particular algebraic function on this surface generates a finite dimensional Hamiltonian system. All zeros of this function generate motion constants.
According to Riemann-Roch theorem, the number of zeros is complete and this system is completely integrable (which is very plausible).
If the Riemann surface is non-compact and consists of infinite amount of sheets, the question of integrability remains open.

\subsection*{References}

\begin{enumerate}

\item A.I. Dyachenko, V.E. Zakharov, Is free-surface hydrodynamics an integrable system?, Phys. Lett. A 190 (2), 144-148 (1994).
    
 \item V.E. Zakharov, A.I. Dyachenko, Free-Surface Hydrodynamics in the conformal variables, arXiv:1206.2046.
     
     \item M.S. Longuet-Higgins, A technique for time-dependent free-surface flow, Proc. R. Soc. Lond. A 371 (1980), 441-451.
         
 \item V.E. Zakharov, Stability of periodic waves of finite amplitude on the surface of a deep fluid, Zh. Prikl. Mekh. Tekh. Fiz.,  9(2), 86-94 (1968); English: J. Appl. Mech. Tech. Phys., 9(2), 190-194 (1968/1972)

 \item A.I. Dyachenko, E.A. Kuznetsov, M.D. Spector, V.E. Zakharov, Analytical description of the free surface dynamics of an ideal fluid (canonical formalism and conformal mapping), Phys. Lett. A 221 (1-2), 73-79 (1996).

 \item A.I. Dyachenko, V.E. Zakharov, E.A. Kuznetsov, Nonlinear dynamics of the free surface of an ideal fluid, Plasma Phys. Repts., 22(10), 829-840 (1996).
   
\item   A.I. Dyachenko, On the dynamics of an ideal fluid with a free surface, Dokl. Math., 63(1), 115-117 (2001).
    
    \end{enumerate}

\end{document}